\documentclass[final,5p,times,twocolumn]{elsarticle}

\usepackage{hyperref}
\usepackage{graphicx}
\usepackage{dcolumn}
\usepackage{bm}

\usepackage{amssymb}

\journal{Computer Physics Communications}

\begin{document}

\begin{frontmatter}



\title{A rigorous sequential update strategy for parallel kinetic Monte Carlo simulation}


\author{Jerome P. Nilmeier}
\ead{nilmeier1@llnl.gov}
\author{Jaime Marian}%
\ead{marian1@llnl.gov}

\date{\today}

\address{
 Physical and Life Sciences Directorate\\
 Lawrence Livermore National Laboratory, 7000 East Avenue, Livermore CA, 94550}

\begin{abstract}
The kinetic Monte Carlo (kMC) method is used in many scientific fields in applications involving rare-event transitions. Due to its discrete stochastic nature, efforts to parallelize kMC approaches often produce unbalanced time evolutions requiring complex implementations to ensure correct statistics.  In the context of parallel kMC, the sequential update technique has shown promise by generating high quality distributions with high relative efficiencies for short-range systems.  In this work, we provide an extension of the sequential update method in a parallel context that rigorously obeys detailed balance, which guarantees exact equilibrium statistics for all parallelization settings.  Our approach also preserves nonequilibrium dynamics with minimal error for many parallelization settings, and can be used to achieve highly precise sampling.
\end{abstract}


\end{frontmatter}


\section{ \label{sec:Introduction} Introduction}

Since its development in the 1970s, the kinetic Monte Carlo (kMC) method~\cite{Bortz197510, gillespie1976general,kalos2008monte} has enjoyed wide popularity, and has been applied to problems far beyond what it was initially designed to model. The kMC approach belongs to a general class of methods known as stochastic \emph{discrete event simulators}~\cite{jefferson1985virtual}, which have also attracted much attention and have been used in numerous applications. These simulation techniques are mesoscale by design, as the inputs are often \emph{propensities} --or probabilities per unit time-- that are extracted from either simulations, measurements, or both, at microscopic scales.  Due to the fact that it is an event driven algorithm, kMC has the potential of vastly extending the accessible timescales of its continuous-time counterparts.

As the demand for simulations of larger system sizes increases, many questions about how best to parallelize these methods remain.  The main difficulty arises from the fact that standard discrete timestep approaches to parallelization of deterministic differential equations~\cite{trotter1959product} and molecular dynamics integrators~\cite{tuckerman1992reversible} are not directly applicable due to the discrete and stochastic nature of time evolution.  Significant progress towards the formulation of a Trotter decomposition has been achieved, however~\cite{arampatzis2012hierarchical}, and the subject remains an active area of research.

By far, the dominant paradigm in parallel kMC is the asynchronous approach, working from the idea that parallel processes are run simulataneously, with intermittent bookkeepping to recover either exact or nearly exact statistics.  The classic set of rigorous and semirigorous approaches proposed by Amar and Shim~\cite{shim2005semirigorous,shim2005rigorous}, and other derivative algorithms~\cite{fichthorn1991theoretical}, remain as the reference for parallel kMC simulators.  These designs can be highly efficient, but the asynchronous strategy gives rise to rough \emph{virtual time horizons}, since each process advances by an independent, stochastically varying time clock.  This effect was first noted and addressed by Korniss {\it et al}~\cite{korniss2000massively,korniss2002statistical,korniss2003suppressing}.  Martinez {\it et al}~\cite{martinez2011billion,martinez2008synchronous} proposed an elegant solution to the time horizon problem, resulting in a synchronous approach.  They build on a null event formulation and that is found in the discrete event community, while also developing a controlled approximation to the master equation for the parallelized process.

In recent years, a number of researchers have developed the notion of \emph{sequential updates} in the context of single process simulations~\cite{ren2006acceleration,orkoulas2009spatial,noon2010simulation,orkoulas2010spatial}.  This formulation has been rigorously shown to produce equilibrium distributions by obeying a balance condition, but has not been applied to parallelization contexts. One advantage to the sequential approach in a parallel simulation is that the time steps advance sequentially with each process, and there is no need to worry about time synchrony across all processes.  A recent sequential approach for parallel simulations has been proposed by Arampatzis, {\it et al}~\cite{arampatzis2012hierarchical}.   Since the sequential method can limit parallel efficiency, considerable care is given as to how to treat noninteracting processes simultaneously, making a convincing case that this approach can also be efficient.  In the SPPARKS simulation suite of Plimpton {\it et al}~\cite{plimpton2009crossing}, an efficient implementation that simultaneously updates noninteracting processes is also used.

In this work, we propose to further develop the sequential update paradigm as a parallelization strategy.  Here, however, we propose a procedure that obeys detailed balance, and also show that we can recover a very good approximation to the time response, laying the foundation for more detailed treatments in the future. The basis of our approach lies in defining a procedure for generating sequential update schedules in such a way that detailed balance is assured for the endpoints of the schedules.  We also discuss how to use the statistics of the endpoints of each process within the schedule.  For the present work, we discuss only the theoretical and algorithmic aspects of the parallel protocol, and will cover implementation and performance related issues in a future work.

This paper is organized as follows. We begin with a theoretical overview of the kMC method and sequential updates. We then formulate the sequential update strategy that preserves detailed balance in the context of parallel simulations.  We include a discussion on collecting statistics after each parallel process has run during a schedule sweep.  The method is then tested and applied to Ising systems of increasing complexity in Section \ref{sec:Results and Discussion}.  We conclude with a brief discussion of the results obtained and the conclusions.

\section{Theory:  Sequential Updates with kinetic Monte Carlo\label{sec:Theory} }

\subsection{The master equation and kinetic Monte Carlo}

To begin the discussion, we express our propagation strategy as a master equation.  For kinetic Monte Carlo, it is sometimes more convenient to work with the Chapman-Kolmogorov form, as it contains the transition kernel explicitly in the expression.  The Chapman-Kolmogorov form~\cite{van1992stochastic} of the master equation for a Markov system is
\begin{equation}
T^Np(\sigma;s)=p( \sigma;s+N),
\label{eq:CK_canonical in s}
\end{equation}
where the transition kernel $T$ is expressed in left stochastic form ~\cite{van1992stochastic}, and $p(\sigma;s)$ is the time dependent probability vector for the configuration vector $\sigma$ at integer time state s.  The system is a Markov process, and we advance the system by $N$ steps by applying the $T$ matrix to $p$ an integer number $N$ times.  For the present work, the vector of configurations $\sigma$ is the vector of all $2^{N_S}$ possible spin states where $N_S$ is the number of spins in the system, as is described in \ref{sec:Ising Model Definitions}.  The standard detailed balance for a single step can be expressed as $\pi_i T_{ji} = \pi_j T_{ij}$, where $\pi_i$ is the equilibrium probability of occupying the $i$-th configuration (spin) state, such that $p(\sigma_i;s=\infty)=\pi_i$.  For the cases presented in this work, each time step follows Glauber dynamics~\cite{glauber1963time}, as defined in Equation (\ref{eq:Glauber_TM}).

The timestep in Equation (\ref{eq:CK_canonical in s}) can be advanced by using a Poisson variate for a procedure consisting of $N$ steps, or
\begin{equation}
p_p(\Delta t(N))=\frac{1}{\tau_S}
                       \frac{ \Delta t^{ N-1 } }{(N-1)!}
                       e^{-\Delta t/\tau_S}.
\label{eq:Poisson Distribution}
\end{equation}
The expectation value of a variate drawn from the distribution in Equation (\ref{eq:Poisson Distribution}) is $\left \langle \Delta t(N)\right \rangle = N\tau_S$, where $\tau_S$ is the time scale of the system.  For purposes of this work, the timestep is advanced only by the average value in order to simplify the analysis.  Both approaches give equivalent statistics, however.  For $n$-fold way simulations \cite{Bortz197510,gillespie1976general}, the timescale is computed as the residence time, or the inverse frequency line $\tau_S=1/R$, where $R$ is the sum of all possible transition rates.  For the present work, we regard the $\tau_S$ as the intrinsic timescale of the simulation.  The Poisson variate for an $N$ step process is readily obtained by computing the negative logarithm of $N$ uniform variates and summing them.  Using either the Poisson variate or the expectation value, we can advance the time clock as
\begin{equation}
T^N p(\sigma;t)=p( \sigma;t+\Delta t(N)),
\label{eq:CK_canonical in t}
\end{equation}
to generate the time dependent solution to the master equation.  For all cases in this work, the time step is advanced by the expectation value.

\subsection{Construction of Sequential Strategy that Obeys Detailed Balance}

Here we develop a procedure based on the work of Deem \textit{et al}~\cite{manousiouthakis1999strict}, and also Orkoulas \textit{et al}~\cite{orkoulas2009spatial,orkoulas2010spatial,noon2010simulation}, who developed a theory for sequential updates and showed that exact equilibrium distributions can be obtained.  The primary motivation for using sequential updates in these works was to accelerate convergence of equilibrium simulations.  For the present work, we wish to develop the sequential update procedure as a parallelization strategy, following ideas introduced by Shim and Amar \cite{shim2005semirigorous} and Arampatzis \textit{et al}~\cite{arampatzis2012hierarchical}.  Since we wish to have a parallelization strategy suitable for studying nonequilibrium and dynamical properties, the goal here is to develop a procedure that preserves the dynamic character of native, unparallelized simulations, rather than to have rapid convergence properties.  Our procedure can be regarded as an advance in that it introduces a sequential update strategy obeying detailed balance in a parallelization context.

\begin{figure}
\centering
\includegraphics{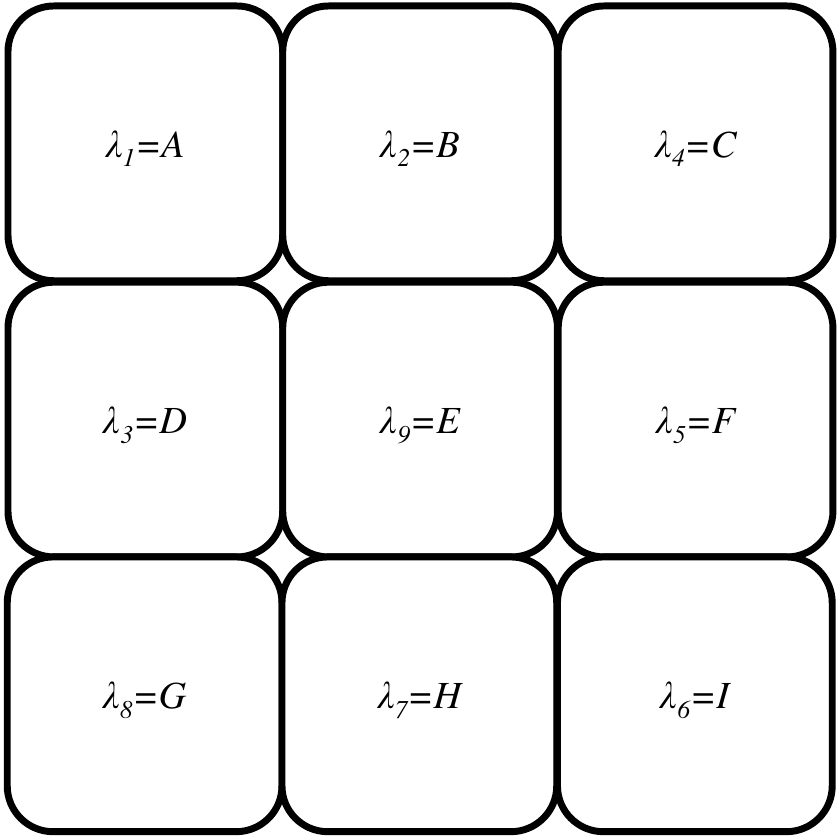}
\caption{\label{fig:replica_diagram} Example 2D configuration partitioned into $N_D=9$ domains with a 'checkerboard' pattern and schedule $\Lambda=\left\{A,B,D,C,F,I,H,G,E\right\}$.  Each domain runs for a fixed number of  $N_I$ independent steps with neighboring (or interacting) domain processes held fixed.  At the end of each sweep, a new schedule is generated randomly.  A vertically striped partitioning would treat the combined domains $(A,B,C)$, $(B,E,H)$, and $(C,F,I)$ each as a single process, resulting in $N_D=3$.}
\end{figure}

Consider a configuration space that is partitioned into domains, such as that shown in Figure \ref{fig:replica_diagram}.  Each domain is of equal size, and the domain partitioning is held fixed for the duration of the simulation.  The general procedure of sequential updating requires the simulation of a single domain process for a number $N_I$ of independent time steps while holding the neighboring domains at a fixed coordinate state.  For each process of length $N_I$, data is first collected from the fixed state of the neighboring processes.  For distributed data parallelizations, this data from neighboring processes is sometimes referred to as a \emph{halo}.  For the Ising system, the halo consists only of the spins from adjacent domains that are in direct contact with the domain that is being simulated.  Once this is complete, another domain is selected according to a schedule
\begin{equation}
\Lambda= \big\{\lambda_1,...,\lambda_d,..,\lambda_{N_D}\big\},
\label{eq:schedule}
\end{equation}
where $N_D$ is the number of domains, and $\lambda_d$ is the $d$-th domain in the schedule.  Each $d$-th process collects halo data prior to running for $N_I$ steps.  An example schedule is given in Figure \ref{fig:replica_diagram}.  The update schedule can thus be in any order, so that all domains are visited in one schedule sweep.  A new schedule is generated for the next round of updates.  Under this definition, there are $N_D!$ possible update schedules.

The sequential requirement can limit efficiency in parallel simulations.  For systems whose interaction ranges extend only to neighboring domains, however, the order of updates is not important, and can be carried out simultaneously, as has been noted previously ~\cite{arampatzis2012hierarchical,plimpton2009crossing}.  We note here that the procedure that we present is also valid for systems whose halos, or interaction ranges, could extend beyond neighboring domains, which can affect the choice of domains to be run simultaneously.  In a future work, we will discuss ways of decomposing short and long range interactions in an efficient way.  For clarity, we develop this procedure for the Ising model, which is given in detail in \ref{sec:Ising Model Definitions}, although the procedure is perfectly general for both discrete and continuous systems.  For the present discussion, the unpartitioned Ising model results in a vector of $2^{N_S}$ discrete states, which  uses a transition kernel $T$ of size $2^{N_S}\times2^{N_S}$.  After applying a partitioning procedure, we require that each domain process obeys detailed balance
\begin{equation}
[T_{(\lambda_d)}]_{nm}=\frac{\pi_n}{\pi_m}
                       [T_{(\lambda_d)}]_{mn},
\label{eq:replica detailed balance}
\end{equation}
which also implies that the domain process run for $N_I$ independent steps obeys detailed balance \cite{van1992stochastic}
\begin{equation}
[T_{(\lambda_d)}^{N_I}]_{nm}=\frac{\pi_n}{\pi_m}
                       [T_{(\lambda_d)}^{N_I}]_{mn}.
\label{eq:replica detailed balance_ni}
\end{equation}
Equation \ref{eq:replica detailed balance_ni} can be readily derived for a discrete system by computing the transition matrix product $N_I$ times and taking the $(m,n)$th element of the product.  To derive the detailed balance condition, we first develop a two domain procedure ($N_D=2$).  From the two domain procedure, a three domain procedure is derived, followed by an arbitrary $N_D$ number of domains.  For the current procedure, the partitions are of equal size, and $N_I$ is the same for all domains.

Consider first the procedure where the full configuration space is partitioned into two domains $A$ and $B$.  For the Ising model, this means that only spins assigned to each domain are allowed to fluctuate during the process.  The resulting partitioned transition matrices are expressed as $T_{(A)}$ and $T_{(B)}$.  Although the partitioning allows for these matrices to be rewritten as $2^{N_S/N_D}\times2^{N_S/N_D}$ matrices, we consider the partitioned forms as $2^{N_S} \times 2^{N_S}$ matrices, such that the full reference transition matrix is he superposition $T_{REF}=1/2\cdot( T_{(A)}+T_{(B)} )$.  A full sweep across domains with the schedule $\Lambda=\left\{A,B\right\}$, where each domain process runs independently for $N_I$ steps, would therefore be expressed as $T_{(B)}^{N_I} T_{(A)}^{N_I}p(\sigma;s)$.  The matrix product can be expressed as
\begin{eqnarray}
\left[T_{(B)}^{N_I} T_{(A)}^{N_I}\right]_{JI}
&=&\sum_k \left[T_{(B)}^{N_I} \right ]_{Jk}\left[T_{(A)}^{N_I} \right ]_{kI}
\nonumber\\
&=&\sum_k \left[\frac{\pi_J}{\pi_k}T_{(B)}^{N_I} \right ]_{kJ}
            \left[\frac{\pi_k}{\pi_I}T_{(A)}^{N_I} \right ]_{Ik}
\nonumber\\
&=&\frac{\pi_J}{\pi_I}\left[T_{(A)}^{N_I}T_{(B)}^{N_I}\right]_{IJ},
\label{eq:two replica detailed balance}
\end{eqnarray}
which indicates that the forward schedule $\Lambda=\left\{A,B\right\}$ obeys detailed balance with respect to the reverse schedule, $\tilde{\Lambda}=\left\{B,A\right\}$.  Since Equation (\ref{eq:two replica detailed balance}) is for a full full schedule sweep, the endpoints $I$ and $J$ are given in capital letters for clarity.  The $d$-th domain process in the forward schedule is related to the reverse domain process as $\lambda_{d}=\tilde{\lambda}_{N_D+1-d}$.  Consider next the $N_D=3$ schedule $\Lambda=\left\{A,B,C\right\}$.  The transition matrix for this forward sequence is given as
\begin{eqnarray}
\left[T_{(C)}^{N_I}T_{(B)}^{N_I}T_{(A)}^{N_I}\right]_{JI}
&=&\sum_k \left[T_{(C)}^{N_I} \right ]_{Jk}\left[T_{(B)}^{N_I}T_{(A)}^{N_I} \right ]_{kI}
\nonumber\\
&=&\sum_k \frac{\pi_J}{\pi_k}  \left[T_{(C)}^{N_I} \right ]_{kJ} \frac{\pi_k}{\pi_I} \left[T_{(A)}^{N_I}T_{(B)}^{N_I} \right]_{Ik}
\nonumber\\
&=&\frac{\pi_J}{\pi_I}\left[T_{(A)}^{N_I}T_{(B)}^{N_I}T_{(C)}^{N_I}\right]_{IJ},
\label{eq:three replica detailed balance}
\end{eqnarray}
where the result of Equation (\ref{eq:two replica detailed balance}) was used in the second line of Equation (\ref{eq:three replica detailed balance}).  For a sequence with $N_D$ domains, it is straightforward to generalize Equation \ref{eq:three replica detailed balance} to
\begin{eqnarray}
T_{SEQ}(\Lambda;N_I,N_D)&=&\prod _{d=1}^{N_D}T_{(\lambda_d)}^{N_I}\nonumber\\
T_{SEQ}(\tilde{\Lambda};N_I,N_D)&=&\prod _{d=1}^{N_D}T_{\tilde{(\lambda_d)}}^{N_I}=\prod _{d=N_D}^{1}T_{(\lambda_d)}^{N_I}, \nonumber\\
\label{eq:replica detailed balance of sequential tm}\end{eqnarray}
where the number of domains $N_D$ and number of steps $N_I$ run on each domain are parameters of the simulation, while the schedule $\Lambda$ can vary with each sweep.

Using the notation introduced in Equation (\ref{eq:replica detailed balance of sequential tm}), the detailed balance condition for the entire sweep in terms of a particular schedule is
\begin{equation}
\left[T_{SEQ}(\Lambda) \right ]_{JI}=\frac{\pi_J}{\pi_I} \left[T_{SEQ}(\tilde{\Lambda}) \right]_{IJ}.
\label{eq:general reversibility condition}
\end{equation}
For a system where the schedules are selected with some weighted probability $w(\Lambda)$, we simply require that $w(\Lambda_i)=w(\tilde{\Lambda}_i)$ in order to maintain detailed balance.  The easiest way to meet this requirement is to select a schedule randomly at each sweep, so that $w(\Lambda_i)=1/N_D!$.  The resulting transition matrix is a weighted superposition of these transition matrices over all $N_D!$ possibilities
\begin{equation}
T_{PAR}=\sum_{i=1}^{N_D!} w(\Lambda_i)T_{SEQ}(\Lambda_i),
\label{eq:sum of sequential tms}
\end{equation}
The resulting master equation for this system will contain a transition matrix with the correct weighted sum of all $N_D!$ schedules.  If only a subset of schedules is chosen, detailed balance can be maintained as long as the corresponding reverse schedule is included.  For all cases studied here, we selected schedules uniformly.  It should be noted that $T_{PAR}(N_I,N_D) \neq T_{
REF}^{(N_I \cdot N_D)}$ in general, but they are often sufficiently similar that simple scaling of dynamics can reproduce the nonequilibrium response, as is shown in section \ref{sec:Results and Discussion}.

Equation (\ref{eq:CK_canonical in s}) can be rewritten for a parallel system as
\begin{equation}
T_{PAR} \cdot p(\sigma;s)=p( \sigma;s+N_I \cdot N_D),
\label{eq:CK_parallel in s}
\end{equation}
where we note that the intermediate timesteps are also accounted for in the integer timestep.

\subsection{Collecting Statistics and Time Scaling\label{subsec: Stats and Time Scaling}}

Equation (\ref{eq:general reversibility condition}) demonstrates that only the endpoints of the sweep obey detailed balance, and makes no claim as to whether the intermediate steps can be used to gather statistics.  In general, the intermediate steps of a general Metropolis-Hastings procedure cannot be used ~\cite{hetenyi2002multiple,nilmeier2008multiscale,nilmeier2011nonequilibrium},
unless a reweighting scheme is applied.  For the procedure presented here, this limitation holds for the independent steps on each process.

We can, however, use the statistics generated at the endpoint of each domain process run for $N_I$ steps.  Consider a sequence of three randomly generated schedules $...\left \{C,A,B \right \}\left \{b,a,c\right \}\left \{A,B,C\right \}...$, focusing on the middle schedule, $\left \{b,a,c\right \}$.  According to strict interpretation of the present protocol, only the data after domain process $c$ could be collected.  We can relax our definition of schedule to include duplicates.  Since the entire schedule is rejection free, we can treat the register shifted sequences as valid sources of statistics.  The same sequence can be seen to consist of the schedules $\left \{...,C \right \}\left \{A,B,b \right \}\left \{a,c,A \right \}\left \{B,C,...\right \}$, and $\left \{...C,A\right \}\left \{B,b,a\right \}\left \{c,A,B\right \}\left \{C,...\right \}$.  While the original schedule has $c$ as the endpoint, the second and third sequences contain $b$ and $a$ as the endpoints, respectively.  It should be noted, however, that we cannot make a similar assertion about the intermediate statistics generated during the domain process.  This can be clearly seen in Figure \ref{fig:ni example}, and is discussed in detail in section \ref{subsec: 1D Results}.

To add the time scale, we assert that each simulation procedure has a single decay constant that can be measured by computing the equilibrium autocorrelation function $C_A(\Delta s)=\left\langle A(s)A(s+\Delta s)\right\rangle$ of some observation $A$ and finding a function $C_A(\Delta s;\tau_{PAR}) =C_A(0) e^{-\Delta s/\tau_{PAR}}$.   For all cases studied here, $A(s)=H(s)m(s)$, where $H(s)$ and $m(s)$ are the energy and magnetization as defined in \ref{sec:Ising Model Definitions}.  The autocorrelation function is computed for each trajectory, and the average of these functions over $N_{TRAJ}$ realizations is used to compute the autocorrelation function from which $\tau_{PAR}$ is extracted.  The timestep of any parallelized simulation can then be rescaled to the timescale of the standard simulation by using $t= s \cdot \tau_{REF}/\tau_{PAR}$.

\section{Results and Discussion\label{sec:Results and Discussion}}

\subsection{1 dimensional Ising model\label{subsec: 1D Results}}

A 1D Ising model with no external field and toroidal boundary conditions was studied using the potential given in Equation (\ref{eq:Hamiltonian Definition}).  As is the case with all simulations studied, Glauber dynamics are used, as is given in \ref{eq:Glauber_TM}.  The number of spins is $N_S=8$, resulting in a $2^{N_S}=256$ states.  We prepare the system in the $8$ initial conditions corresponding to two consecutive up spins, with the remaining spins in the down state, given as $\sigma_{IC}= \left\{(\downarrow\downarrow\downarrow\downarrow\downarrow\downarrow\uparrow\uparrow),
(\downarrow\downarrow\downarrow\downarrow\downarrow\uparrow\uparrow\downarrow),...
(\uparrow\uparrow\downarrow\downarrow\downarrow\downarrow\downarrow\downarrow),
( \uparrow \downarrow\downarrow\downarrow\downarrow\downarrow \downarrow \uparrow)\right\}$.  These initial conditions are identical for the unpartitioned case, since the boundaries are symmetric.  The location of the partitions break this symmetry however, which is why we chose to study different locations of the paired spin up condition.  Using the first initial condition as an example, the partitions $N_D=\left\{1,2,4,8\right\}$ can be shown as
\begin{eqnarray}
N_D&=&1: (\downarrow\downarrow\downarrow\downarrow\downarrow\downarrow\uparrow\uparrow)\nonumber\\
N_D&=&2: (\downarrow\downarrow\downarrow\downarrow)(\downarrow\downarrow\uparrow\uparrow)\nonumber\\
N_D&=&4: (\downarrow\downarrow)(\downarrow\downarrow)(\downarrow\downarrow)(\uparrow\uparrow)\nonumber\\
N_D&=&8: (\downarrow)(\downarrow)(\downarrow)(\downarrow)(\downarrow)(\downarrow)(\uparrow)(\uparrow ),
\label{eq:1D partitions}
\end{eqnarray}
where each set of spins in parentheses are treated as a separate domain, and is run for $N_I$ independent steps.  We use four different settings for $N_I$, chosen to be  $N_I=\left\{1,2,5,10\right\}$.  Note that the introduction of a partitioning scheme in general is expected to produce dynamic artifacts which are studied here empirically.  In order to gather good statistics in both the nonequilibrium and equilibrium regimes, multiple trajectories $N_{TRAJ}=6000$ are collected (see Table \ref{tab:table1}) for each setting of $(\sigma_{IC},N_I,N_D)$.  A new trajectory for each setting is generated by generating a new random seed prior to running the simulation.  The dimensionless temperature is set to $\beta J=0.5$, and each each trajectory is run for a total of $S=800$ steps, regardless of parallelization settings.  Statistics are gathered after each number of $N_I$ steps are run.  Time dependent observations are computed as
\begin{equation}
\left \langle A(s) \right \rangle =\frac{1}{N_{TRAJ}}\sum_{i=1}^{N_{TRAJ}}A_i(s)
\label{eq:time dependent averages}
\end{equation}
where $A_i(s)$ is value of observation $A$ at timestep $s$ for the $i$-th trajectory. For the 1D cases, these trajectories can include either a set of trajectories from a particular initial condition, or as a sum over multiple realizations of multiple initial conditions.
\begin{figure}
\begin{center}
\includegraphics{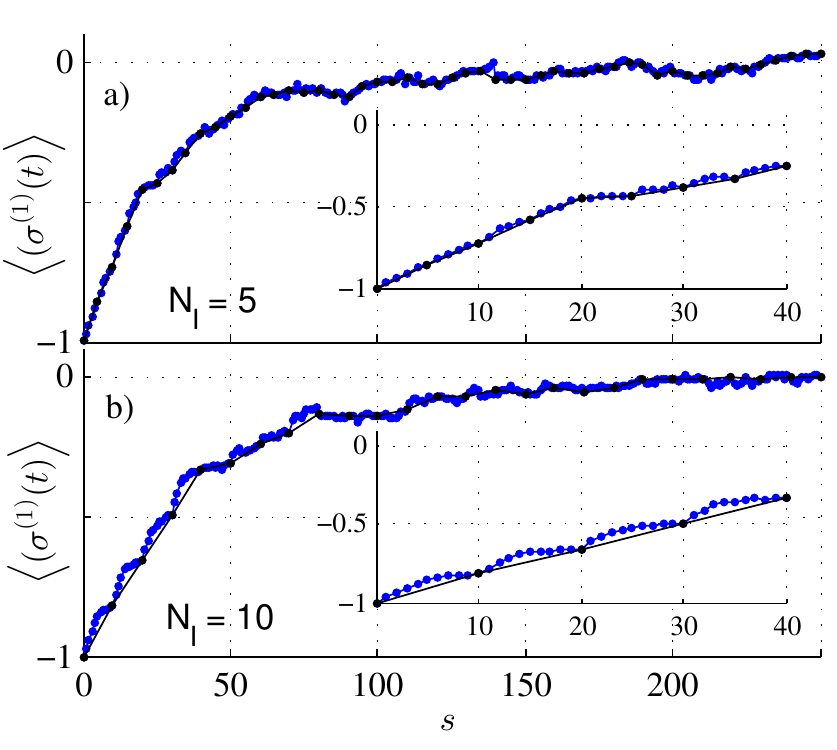}
\caption{ \label{fig:ni example}
Timeseries examples from 1$\times$8 spin Ising model, $N_D=4$, $\sigma_{IC}=( \downarrow\downarrow)(\uparrow\uparrow)(\downarrow\downarrow)(\downarrow \downarrow )$. Data is also shown in Figure \ref{fig:spin-average-timeseries}. a) Time series for average of spin 1, with $N_I=5$ and b) $N_I=10$.  Data points in black are collected for statistics, while the intermediate steps are discarded.  Inset panels on right hand side of figures are the first 40 steps of the same dataset.
}
\end{center}
\end{figure}

The intrinsic time constant $\tau_{PAR}$ is extracted from the autocorrelation function as mentioned in Section \ref{subsec: Stats and Time Scaling}.  The product of energy and magnetization was chosen to break the degeneracy of energies that can result from computing configurations that are equal in energy but not easily interchangeable through simulation steps.  This effect is particularly important for very small systems, whose energies equilibrate very rapidly, which will be discussed in Section \ref{subsec:Time Response Discussion}.  Once the $\tau_{PAR}(N_I,N_D)$ values are computed, all observations for all parallelization schemes can be plotted on the same axis scaled by $s/\tau_{PAR}$.

As can be seen from Fig. \ref{fig:ni example}, the statistics for the intermediate steps show a distinct response behavior, and are discarded.  This effect becomes more pronounced with fewer spins per domain, as well as increased $N_I$.  The decay behavior can be seen as a rapid equilibration of each domain to the nonequilibrium boundary conditions of the adjacent domains.  In general, this intermediate effect propagates to an overall distortion of the time responses.  For those responses whose time constants are much larger than $N_I \cdot N_D$, this effect is less noticeable, as in Figure \ref{fig:ni example}a.

\begin{figure}[ht]
\begin{center}
\includegraphics{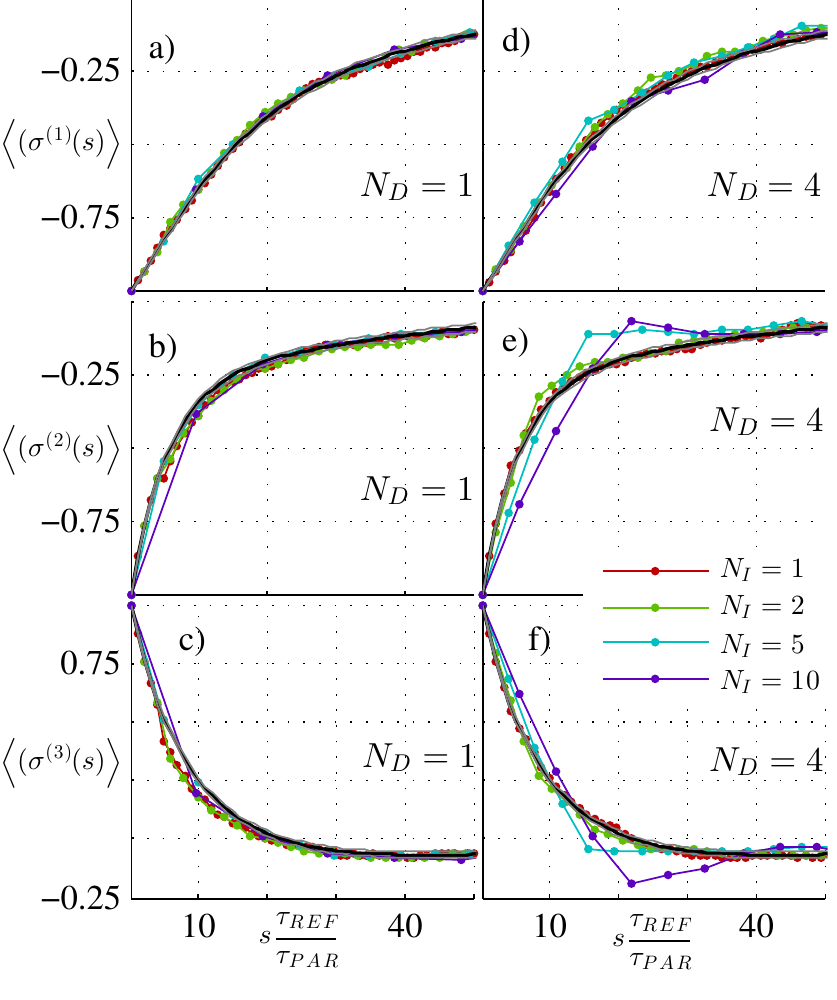}
\caption{\label{fig:spin-average-timeseries}
Typical response behavior for partitioned 1D Ising models.  $\sigma_{IC}=( \downarrow\downarrow\uparrow\uparrow\downarrow\downarrow\downarrow\downarrow)$, at different number of independent steps $N_I=\left\{1,2,5,10\right\}$.  a), b), and c) show spin average time series for spins 1,2, and 3 with $N_D=1$ (no partition, but interval statistics are used).  d), e) and f) show spin averages for spins 1, 2, and 3 with $N_D=4$.
}
\end{center}
\end{figure}

The 1D systems were chosen so that details of the configuration evolution could be studied.  In Figure \ref{fig:spin-average-timeseries}, the spin averages $\left \langle \sigma^{(i)}(s)\right \rangle $ where $i$ is the spin index, are computed as time dependent averages using Equation (\ref{eq:time dependent averages}).  Spins 1, 2, and 3 are plotted for an unpartitioned system starting from the same configuration and different partitioning schemes.  The unpartitioned system ($N_D=1$) shows artifacts resulting only from taking only the $N_I$-th data point.  Some of the fast time response behavior for this system happens faster than $N_I$, and this artifact can in fact introduce a numerical error for such a small system.  For example, the energy $\left \langle H(s)\right \rangle $ passes through a local maximum at $s\approx5$, causing difficulties in systems of $N_I=5$ and $N_I=10$, regardless of partitioning scheme.  See Section \ref{subsec:Time Response Discussion} for more discussion.  For the $N_D=4$ partitioning of Figure \ref{fig:spin-average-timeseries}, notable distortions can be seen. It is also noteworthy that the distortions in the time response are not homogeneous.  The timeseries for spin 1, for example, appears to scale very well, while spins 2 and 3 show a response which rapidly approaches a point that overshoots the actual response, and gradually relaxes to the longer time decay.  This happens at an interface between spins, and also a domain boundary.  Each domain process is experiencing a frozen boundary condition for $N_I$ steps, which distorts the response during that sweep.  This effect can be particular to the location of each spin relative to the partition, and a single scaling constant may not always capture this more complex behavior.

The equilibrium behavior of these systems is guaranteed by Equations (\ref{eq:general reversibility condition}) and (\ref{eq:sum of sequential tms}).  In fact, the equilibrated data are far better, because we can invoke the ergodic principle and obtain multiple estimates of the equilibrated state from a single trajectory in the equilibrium regime, in addition to those data from the multiple trajectories.  We report only the simulated estimate and compare to the generated by evolving the probability distribution to equilibrium according to Equation (\ref{eq:CK_parallel in s}) in Table \ref{tab:table1}.  All other settings match this to within the precision reported for the unpartitioned case.

%

\begin{table*}
\begin{center}
\caption{\label{tab:table1} Settings and statistics for unpartitioned simulations. }

\begin{tabular}{|c|rrrrrr|}
\hline
 &\multicolumn{3}{c}{Settings}
 &\multicolumn{2}{c}{Simulation Estimate}
 &\multicolumn{1}{c}{Analytic}\vline\\

 Spins & $1/\beta J$ & $S$ & $N_{TRAJ}$ & $\tau_{REF}$ & $\left \langle  H\right \rangle  /N_S$ & $\left \langle H\right \rangle /N_S$ \\ \hline
1$\times$8     & 2 & 800           & 6000        & 31.80 &
         $-0.237(0.138)$ & $-0.238(0.138)$ \\
12$\times$12   &2.4 & $1.6\times10^5$ & 500        &$1.13\times10^3$ &
        $-1.286(0.235)$  & $-1.239(0.221)$ \\
100$\times$100 &2.4 & $1.6\times10^7$ & 50       & $1.49\times10^6$ &
        $-1.205(0.021)$  & $-1.204(0.026)$ \\
        \hline
\end{tabular}
\end{center}
\end{table*}

\subsection{2-dimensional Ising model\label{subsec: 1D Results}}

A 2D Ising model was used, following Equations (\ref{eq:Hamiltonian Definition}) and (\ref{eq:Glauber_TM}).  We studied two square lattices with $N_S=12^2$ and $N_S=100^2$.  Only one initial condition was used for each system size.  The initial condition was a filled circle.  The radius of the circle was set as $0.7975\sqrt{N_S}$, and spins inside this radius are assigned spin up, while the remaining grid points are set as spin down.  The resulting initial conditions have a magnetization of 0.0056 and 0 for the $N_S=12^2$ and $N_S=100^2$, respectively.  Domains were generated in either a checkerboard pattern or a vertically striped pattern.  For the 12$\times$12 system, we studied $N_D=\left\{2,4\right\}$ for the striped configuration and $N_D=4$ for the checkerboard configuration.  For the 100$\times$100 system, we studied the same partitions as in the 12$\times$12 set, and added another checkerboard configuration with $N_D=16$.  As in the 1D case, we studied $N_I=\left\{1,2,5,10\right\}$ for both 2D systems.  Additional simulation settings are listed in Table \ref{tab:table1}.  A comparative discussion of the timeseries behavior is given in the next Section.

Equilibrium statistics for the 2D systems were also obtained.  The unpartitioned  simulation data are listed in Table \ref{tab:table1}, along with the analytical result using the formula given by Fisher \cite{ferdinand1969bounded}.  The equilibrium histograms are not reported here, but were verified against the 100$\times$100 dataset given in Ref.~\cite{ren2006acceleration}, where we note that the histograms match so well that it is difficult to see any difference at all.  The long time behavior shown in Figure \ref{fig:energy-timeseries} also shows this to be the case.

\subsection{Time Response Behavior \label{subsec:Time Response Discussion}}

\begin{figure}[hbt]
\begin{center}
\includegraphics{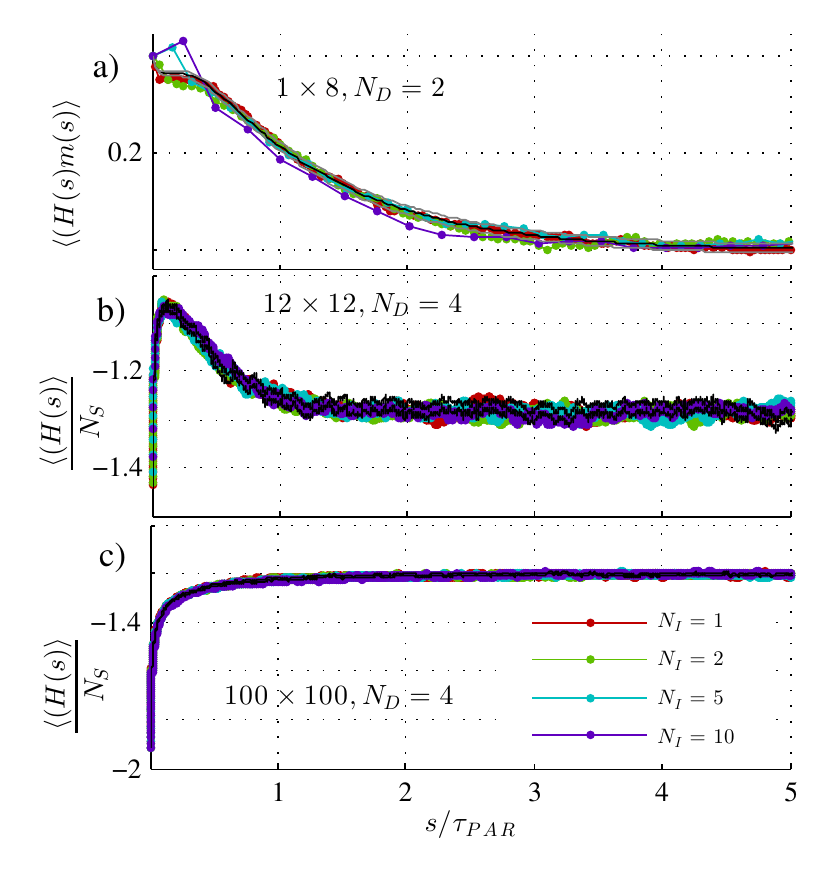}
\caption{\label{fig:energy-timeseries}
Typical timeseries for over $S=5\tau_{PAR}$ used for $\chi^2$ calculations.  a) shows $\left \langle  A(s)\right\rangle  =\left \langle  H(s)m(s)\right\rangle $ for $\sigma_{IC}=(\downarrow\downarrow\downarrow\uparrow)(\uparrow\downarrow\downarrow \downarrow )$ ($N_D=2$).  b) and c) show time response for initial condition as described in text, with $N_D=4$, and checkerboard partitioning.  Data shown for $N_I=\left\{1,2,5,10\right\}$.
}
\end{center}
\end{figure}

\begin{figure*}[hbt]
\begin{center}
\includegraphics{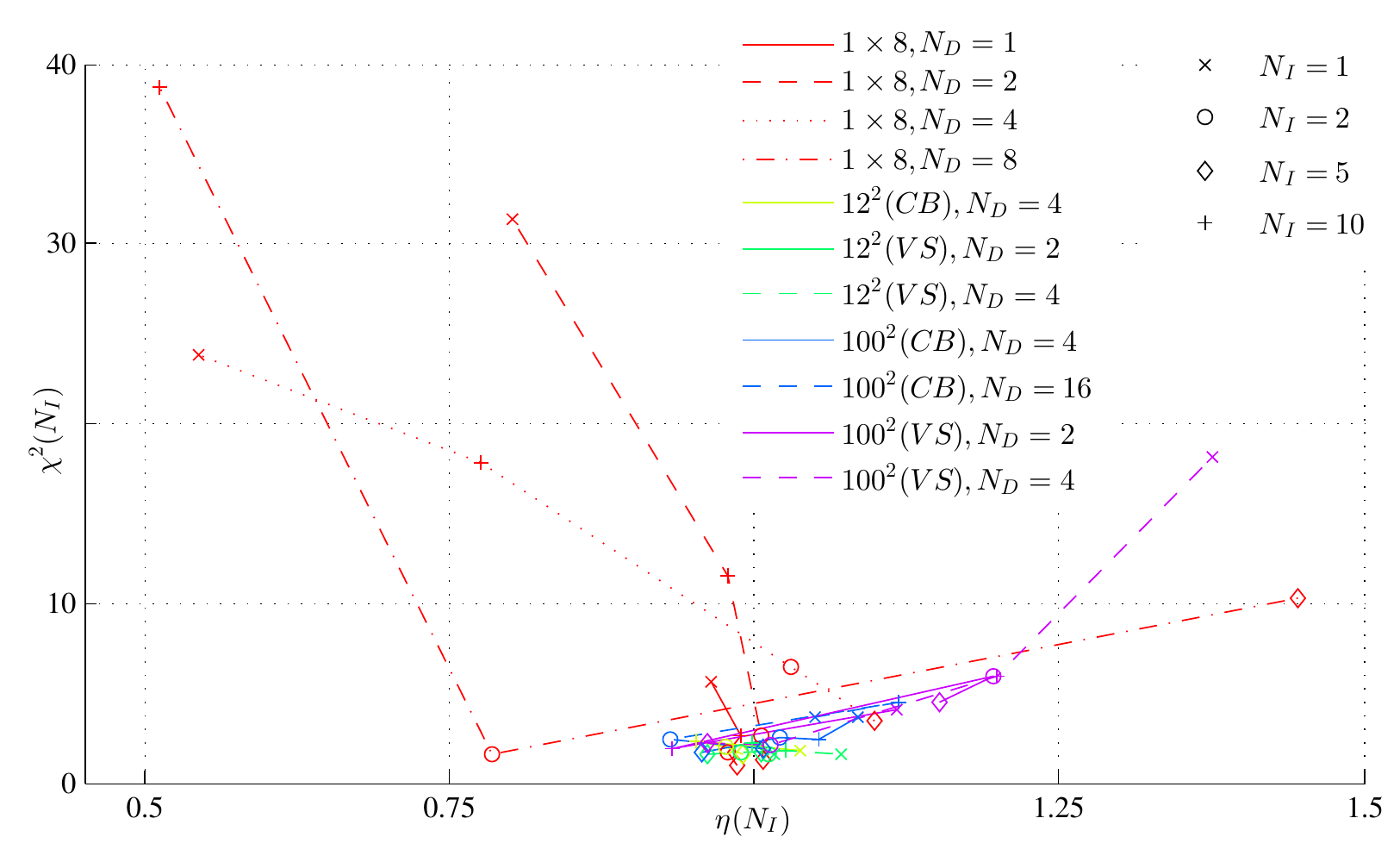}
\caption{ \label{fig:chi_v_eta}
Response Error (Eq. \ref{eq:error metric}) versus Ideal Efficiency (Eq. \ref{eq:efficiency}).  Markers are assigned according to $N_I$ as shown in legend, and colors are assigned according to Ising Model size and $N_D$ as shown in legend.  In general, ideal efficiencies near one correspond to low errors in nonequilibrium response.
}
\end{center}
\end{figure*}

Figure \ref{fig:energy-timeseries} shows example time series from each of the 1$\times$8, 12$\times$12, and 100$\times$100 datasets.  These timeseries were used to compute the $\chi^2$ metric, given in Equation (\ref{eq:error metric}).  The 1D data set has a fast decay phase, where $\left \langle H(s)\right \rangle $ goes through a maximum near $s=5$, which introduces considerable error, while the magnetization function decays very slowly and homogeneously.  The product of these observations was thus selected, in order to capture both the fast and slow responses.  The 12$\times$12 data also go through a local maximum.  This maximum occurs at $s\approx1400$, which is far greater than the largest $N_I$ under study.  The correlation time of $1.13\times10^4$ is also much greater than any interval effect.  As a result, even though there are multiple decay modes apparent in the timeseries, a single scaling of the timeseries appears to be adequate to reconstitute the timeseries of parallelized systems.  In this regard, the performance only improves when increasing the system size.  We obtain the added benefit of more monotonic decay as system displays more supercritical behavior, and note that many larger values of $N_I$ could easily be used for this system.

The performance of each system studied is given in Figures \ref{fig:chi_v_eta} and  \ref{fig:full-summary}.  The ideal efficiency $\eta$, as defined in Equation (\ref{eq:efficiency}), gives a rough idea as to the convergence rates of the parallelized systems versus the unparallelized systems.  Here, the ideal efficiency measures ratios of convergence rates for parallel systems that use the same number of timesteps, or energy function calls.

While it is true that increased convergence may be a desirable property of a parallelized system, Figure \ref{fig:chi_v_eta} shows that  that deviations from unity in either direction are strongly correlated with larger response errors.  As a result, an ideal parallel efficiency of $\eta\approx1$ is most indicative of a faithful reconstitution of the unpartitioned dynamics.  As might be expected, increasing $N_I$ tends to also increase the error.  Figure \ref{fig:chi_v_eta} also shows that most cases studied are well behaved, with $\eta \approx 1$ and $\chi^2\approx2$.

In Figure \ref{fig:full-summary}, we see that the 1$\times$8 system shows both upward and downward trends in efficiency with increasing $N_D$ for $N_I=1$ and $N_I=2$, which may be indicative of the quality of the estimate of the correlation times.  A trend is noticeable for $N_I=5$ and $N_I=10$, which can be interpreted.  The extreme case of $N_D=8$ has one spin per domain, which converges to a local equilibrium within $N_I=1$ iteration.  While the domain is in equilibrium with the neighboring boundary conditions, the system is not globally equilibrated.  Therefore, the progress of each domain process is not improved with increasing $N_I$ beyond 1, and so the remaining independent steps are not productive.  The same effect is observed for $N_D=4$.  For these overly parallelized systems, we also observe significant error in the timeseries, for reasons that were also described in Section \ref{subsec: 1D Results}.  The error is most notable in the fast decay region, where the energy portion of the observation goes through a rapid maximum, while the longer decay behavior seems to be better preserved.

For the 2D data, most of the error and efficiency measures are stable and well behaved. The only noticeable trend is the slight increase in efficiency of the vertical stripe partitions with $N_D$, along with an increase in error.  The vertical partition was chosen because it breaks the symmetry of the circular initial condition, and it was believed that this may lead to artifacts that are not as apparent when using a more symmetric (checkerboard) partition.  Indeed, this appears to be the case, and the intuition that symmetric parallelization will lead to better statistics is supported for the cases studied here.

\section{Conclusions}

 We have presented a sequential parallelization protocol that is rejection free and guarantees equilibrium statistics for all parallelization settings.  Additionally, we have applied a simple scaling law that allows the dynamics of one parallelization scheme to be directly compared to an unpartitioned simulation.  

 Some limitations to the method include the requirement to compute the time constant from the autocorrelation time.  The need to account for multiple decay modes in the autocorrelation function while still using a single decay constant as the time scale can introduce error.  Another limiting feature of the autocorrelation time approach is that it requires an equilibrium timeseries.  For systems that are subjected to frequent external time varying fields, this can be cumbersome, and an approach where the intrinsic scaling is computed \textit{in situ} would be more useful.  We intend to build on the previously described approach that uses the frequency line formulation to compute time constants.  Additionally, a method that more properly accounts for the distortion of the fast time dynamics would be most useful for a robust algorithm.  We hope to address these issues in a future work, along with more careful estimates of the error and methods for accounting for long range interactions.

 The current method works very well for reasonably large homogenous systems, however, and we found that that symmetric domain partitioning can better maintain the integrity of the parallel simulations.  Many of the artifacts observed in the 1D systems will be completely avoided, given that partitioning is usually driven by memory requirements, rather than some other motivation that will drive $N_D$ to be unduly large.  The dynamics will also be preserved as long as $N_I<<\tau_{PAR}$, which is a reasonable limitation, given that one would wish to print out and observe dynamics with similar considerations.  Since writing data takes much longer than parallel communication, choosing $N_I$ to coincide with a practical printing frequency will minimize most artifacts of parallelization seen with the present scaling approach.

\section*{Acknowledgments}
This work has been supported by JM's DOE Early Career Research Award.  Computations were carried out on Livermore Computing cluster ansel.  This work performed under the auspices of the U.S. Department of Energy by Lawrence Livermore National Laboratory under Contract DE-AC52-07NA27344.   Release: (LLNL-JRNL-651809-DRAFT).  JM and JN would like to thank Malvin Kalos, Tomas Oppelstrup, Vasily Bulatov, and Eric Darve for helpful comments and critical reading of the manuscript.
%

\appendix

\section{Ising Model: Energy Function and Glauber Dynamics\label{sec:Ising Model Definitions}}

For the present work, we study only the 1D and 2D Ising models.  The 1D model has toroidal periodicity, and the 2D model is periodic in both dimensions, as is standard practice.  The total energy for these systems is given explicitly for both 1D and 2D systems as

\begin{eqnarray}
\beta H(\sigma_j;1D)=&-\beta J\sum_{n=1}^{N_S} \sigma_{j}^{(n)} \sigma_{j}^{(n+1)}
\nonumber\\
\beta H(\sigma_j;2D)=&-\beta J\sum_{m=1}^{\sqrt{N_S}}\sum_{n=1}^{\sqrt{N_S}} \sigma_{j}^{(m,n)} \sigma_{i}^{(m,n+1)} \nonumber\\
&+ \sigma_{j}^{(m,n)} \sigma_{i}^{(m+1,n)},
\label{eq:Hamiltonian Definition}
\end{eqnarray}
where $J$ is the coupling constant, $\beta=1/k_B T$, and $\sigma$ is a vector of configurations defining the state of the system.  For both systems, $\sigma$ is a vector of length $2^{N_S}$, where $N_S$ is the number of spins in the system.  For the 1D system, the $n$-th spin of the $j$-th configuration is notated as $\sigma^{(n)}_j$.  For the $2D$ sum, array notation is used, where the $(m,n)^{\rm th}$ row and column of the $j^{\rm th}$ configuration indicated as $\sigma^{(m,n)}_j$.  For both cases, each spin value can take on values of either 1 or $-1$.
The boundary conditions are periodic, so that the last summand includes the last and first spins of each row. $J$ is the coupling constant, and $\beta=1/{k_B T}$ is the inverse temperature, such that $\beta J$ is dimensionless.  The magnetization is defined as the average of all spins for a given time configuration.  The (unnormalized) equilibrium distribution is
\begin{equation}
\pi_j=e^{-\beta H(\sigma_j)}
\label{eq:equilibrium distribution},
\end{equation}
and the Glauber dynamic \cite{glauber1963time} is computed as
\begin{eqnarray}
T^{(G)}_{ji}=\frac{N_D}{N_S}\frac{\pi_j}{(\pi_i+\pi_j)}
\nonumber\\
T^{(G)}_{ii} = 1-\sum_{i \neq j} T^{(G)}_{ij}.
\label{eq:Glauber_TM}
\end{eqnarray}
where the $N_D/N_S$ is the selection probability for a single spin within a system partitioned into $N_D$ domains.  The Glauber dynamic is usually expressed in a form more amenable to closed form solutions, and includes hyperbolic tangent terms.  Here, we present it an exactly equivalent form, sometimes referred to as the Barker acceptance criterion~\cite{barker1965monte}.

\section{Fitness Metrics \label{sec:Fitness Metrics}}

\begin{figure}[htb]
\begin{center}
\includegraphics{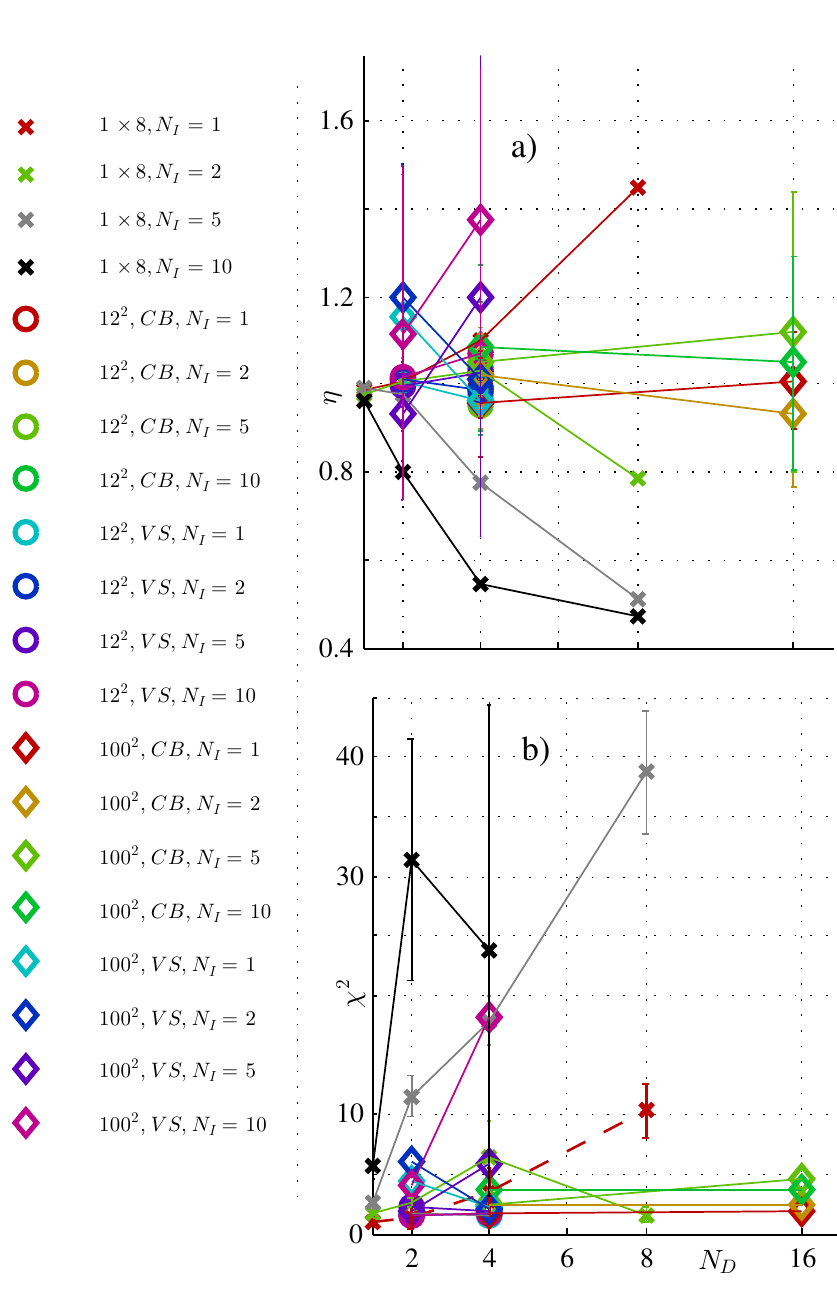}
\caption{\label{fig:full-summary}
Summary of Parallel Performance for all Ising systems studied.  a) Ideal efficiency calculation as defined by Equation (\ref{eq:efficiency}).  b) Response error, as computed in Equation (\ref{eq:error metric}), with example timeseries shown in Figure \ref{fig:energy-timeseries}.
}
\end{center}
\end{figure}

The two fitness metrics used are an efficiency measure $\eta$ and an error measure $\chi^2$.  The first measure is the ideal scaling efficiency, given as
\begin{equation}
\eta(N_I,N_D)=\frac{\tau_{REF}}{\tau_{PAR}(N_I,N_D) \cdot N_I},
\label{eq:efficiency}
\end{equation}
and is similar to what is typically referred to as strong scaling efficiency.  Since we are not carrying out a truly parallelized simulation, with interprocessor overhead, we use the 'ideal' measure only to compare convergence rates of parallel simulations.

The second metric is the error in the nonequilibrium response.  We know from Equation (\ref{eq:general reversibility condition}), that the equilibrium properties are exact.  The time dependent properties, however, are not guaranteed with the present strategy, and we must develop a measure of the error in the nonequilibrium portion of the trajectory.  The error metric is given as
\begin{equation}
\chi^2=\frac{1}{S}\sum_{s=0}^{S}
\frac{
\left[ \left \langle A^{(REF)}(s) \right \rangle -
       \left  \langle A^{(PAR)}( s')\right \rangle \right ]^2
}{
\left \langle \left [  \delta A^{(REF)}(s)\right ]^2 \right \rangle
},
\label{eq:error metric}
\end{equation}
where $s'=s\cdot\tau_{REF}/\tau_{PAR}$, and takes on spline interpolated values to  coincide with the integer values of the reference time increment $s$.  For the 1D Ising system $A(s)=H(\sigma(s))m(s)$, and $A(s)=H(\sigma(s))$ for the 2D systems.  For all cases presented $S=5\cdot\tau_{REF}$.




\bibliographystyle{elsarticle-num}
\section*{References}
\bibliography{kmc_for_arXiv}





\end{document}